\documentclass[aps,pra,twocolumn,showpacs,floatfix]{revtex4}
\usepackage{graphicx}
\usepackage{nicefrac}
\usepackage{amsmath}
\usepackage{amsfonts}
\usepackage{amssymb}
\usepackage{amsthm}
\usepackage{epsf}
\usepackage{bm}
\usepackage{bbm}
\usepackage{longtable}

\usepackage{dcolumn}

\newcolumntype{.}{D{x}{}{-1}}

\newcommand{\bfx}{\bm{x}}

\newcommand{\Za}{{Z\alpha}}

\newcommand{\vare}{\varepsilon}

\newcommand{\pr}{^{\prime}}

\newcommand{\Dmatrix}[4]{ \left(
          \begin{array}{cc}
            #1  & #2   \\
            #3  & #4   \\
          \end{array}
        \right) }

\newcommand{\intinf}{\int^{\infty}_{-\infty}}

\begin{document}

\title{Nuclear size correction to the Lamb shift of one-electron atoms}

\author{Vladimir A. Yerokhin}
\affiliation{Center for Advanced Studies, St.~Petersburg State
Polytechnical University, Polytekhnicheskaya 29,
St.~Petersburg 195251, Russia}

\begin{abstract}

The nuclear size effect on the one-loop self energy and
vacuum polarization is evaluated for the $1s$, $2s$, $3s$, $2p_{1/2}$,
and $2p_{3/2}$ states of hydrogen-like ions. The calculation is performed to
all orders in the binding nuclear strength parameter $\Za$. Detailed
comparison is made with previous all-order calculations and 
calculations based on the expansion in the
parameter $\Za$. Extrapolation of the all-order numerical results obtained towards
$Z=1$ provides results for the radiative nuclear size effect on the hydrogen 
Lamb shift. 

\end{abstract}


\maketitle

\section*{Introduction}

The distribution of charge of the nucleus influences the Dirac energies 
of atomic systems as well as the quantum
electrodynamical corrections to the energy levels.
This effect, termed as the nuclear size (NS) correction, is important for
comparison of theoretical predictions
with experimental data for the whole range of
the nuclear charges $Z$, from hydrogen ($Z=1$) up to uranium ($Z=92$).
The NS corrections to the one-loop self energy and vacuum polarization
have been previously investigated both within the approach
based on the expansion in the nuclear binding strength parameter $\Za$
\cite{friar:79,hylton:85,pachucki:93:density,milstein:03:fns,jentschura:03:jpa,milstein:04}
(where $\alpha$ is the fine structure constant)
and within the numerical approach that accounts the parameter $\Za$ to all
orders \cite{soff:88:vp,mohr:93:prl}. In the high-$Z$ region, the numerical
all-order approach provides accurate predictions for the NS effect on the
radiative corrections. For lower $Z$, however, the NS effect diminishes and
becomes increasingly more difficult to identify in a numerical calculation. On
the contrary, the $\Za$-expansion approach provides accurate predictions
for the low-$Z$ ions and only the qualitative estimates in the high-$Z$ region.
A quantitative cross-check of the two complementary approaches is not
simple and has not been accomplished up to now.

The NS corrections became of particular interest recently, after the results
of the muonic hydrogen Lamb-shift experiment were announced \cite{pohl:10}. It turned
out that the value for the proton charge radius $r_p$
deduced from the muonic hydrogen
differs by five standard deviations from the
spectroscopic value of $r_p$ derived from the hydrogen atom. This unexplained
disagreement  stimulates the scientific community to double-check all
contributions originating from the nuclear charge distribution, both for the
muonic and normal atoms.

The aim of the present investigation is to perform an accurate numerical
all-order calculation of the NS correction to the one-loop self energy and
vacuum polarization and to make a detailed comparison with the
$\Za$-expansion results available.

The relativistic units ($m$ = $\hbar = c = 1$) and the charge units $\alpha =
e^2/(4\pi)$ are used in this paper.

\section{NS correction to Dirac energy}

The leading-order NS correction to the energy levels of a
hydrogen-like atom is defined
as the difference of the corresponding eigenvalues of the Dirac
equation with the point-Coulomb and the extended-nucleus potentials.
The two most commonly used models of the nuclear charge
distribution are the uniformly charged sphere (``sph'') 
and the two-parameter Fermi (``Fer'') model,
\begin{align} \label{1}
\rho_{\rm sph}(r) = \rho_0\,\theta(R_{\rm sph}-r)\,, \\
\rho_{\rm Fer}(r) = \frac{\rho_0}{1+\exp[(r-c)/a]}\,,
\end{align}
where $R_{\rm sph} = \sqrt{5/3}\, R$ is the radius of the sphere with the
root-mean-square (rms) radius $R$, $c$ and $a$ are the parameters of the Fermi
distribution, and $\rho_0$ are the normalization factors. The parameter $a$
of the Fermi distribution is standardly fixed by $a = 2.3/(4\ln3)$~fm.
For a given value of the rms radius, the
parameter $c$ can be determined by the simple approximate formula
\begin{align} \label{2}
c^2 \approx \frac{5}{3} R^2-\frac{7}{3} a^2 \pi^2\,.
\end{align}
In the calculations performed in this work, 
we will assume that the parameter $c$ of the Fermi
distribution is fixed {\em exactly} by the above formula.

For the uniformly charged sphere model, the Dirac equation can be solved
analytically  \cite{shabaev:93:fns}. In this case, the NS correction to
the Dirac energy is represented in terms of the hypergeometric function.
While the exact expression is rather cumbersome, simple
approximate expressions for the NS correction
obtained in Ref.~\cite{shabaev:93:fns} are highly useful.
For the general model of the nuclear charge distribution, the NS correction
can be easily obtained by numerical solution of the Dirac equation.
Numerical results are conveniently parameterized in
terms of the function ${G}_{\rm N}(\Za,R)$, whose definition is inspired by the
analytic relativistic results
\cite{shabaev:93:fns},
\begin{align} \label{5}
\Delta E_{\rm N}(ns) &\ = \frac{(\Za)^2}{n}\,
  \left( \frac{2\Za R_{\rm sph}}{n}\right)^{2\gamma}\, \frac1{10}\,
    {G}_{N}(\Za,R)\,,
  \\
\Delta E_{\rm N}(np_{1/2}) &\ = \frac{(\Za)^4}{n}\,
  \left( \frac{2\Za R_{\rm sph}}{n}\right)^{2\gamma}\,
 \frac{n^2-1}{40n^2}\,  {G}_{N}(\Za,R)\,,
  \label{5a}
\end{align}
where $\gamma = \sqrt{1-(\Za)^2}$ and $R_{\rm sph}$  is
the radius of the sphere with the
rms radius $R$. The function ${G}_{N}(\Za,R)$ is a slowly varying
function of $\Za$ and $R$ and its numerical values are of order of unity.

The numerical results obtained for the NS correction with the Fermi model
of the nuclear charge distribution are listed in
Table~\ref{tab:fns}. The numerical evaluation was performed by solving the Dirac
equation with help of the RADIAL package \cite{salvat:95:cpc} and,
independently, by using the B-spline finite basis set method
\cite{shabaev:04:DKB}. For calculations in the low-$Z$ region, the
RADIAL package was upgraded 
into the quadruple arithmetics (about 32 digits). 
The values of the rms charge radii used were taken from
the compilation \cite{angeli:04} for all ions except for uranium; the uranium
rms radius was taken from Ref.~\cite{kozhedub:08}.

\section{NS correction to self energy}

The one-loop self-energy contribution to the Lamb shift is given by a matrix
element of the self-energy operator with the mass renormalization part subtracted,
\begin{align}  \label{se1}
\Delta E_{\rm SE} = \bigl< a\bigl|\gamma^0 \widetilde{\Sigma}(\vare_a) \bigr|a\bigr>\,,
\end{align}
where $ \widetilde{\Sigma}(\vare) =  \Sigma(\vare)-\delta m$ and $\delta m$ is
the mass counterterm. The self-energy operator is \cite{mohr:98}
\begin{align} \label{se2}
\Sigma(\vare,\bfx_1,\bfx_2)  = 2i\alpha\gamma^0 \intinf d\omega\, &\,
    D^{\mu\nu}(\omega,x_{12})\,
 \nonumber \\ & \times
      \alpha_{\mu}
         G(\vare-\omega,\bfx_1,\bfx_2) \alpha^{\nu} \,,
\end{align}
where $D^{\mu\nu}$ is the photon propagator,
$G$ is the Dirac-Coulomb Green function, $G(\vare) = (\vare-H)^{-1}$,
$H$ is the Dirac-Coulomb Hamiltonian, and $\alpha^{\mu} =
\gamma^0\gamma^{\mu}$ are the Dirac matrices. The nuclear-size self-energy
(NSE) correction is defined as the difference between the matrix elements
(\ref{se1}) evaluated with the point-Coulomb potential and the potential of
the extended-charge nucleus.

Numerical, all-order (in $\Za$)
evaluation of the one-loop self-energy correction have been extensively
discussed in the literature over past decades
\cite{mohr:74:a,blundell:91:se,mohr:93:prl,jentschura:99:prl,cheng:93,yerokhin:99:pra},
both for the case of the point-Coulomb and extended-nucleus potentials. In
the present investigation, we employ the method developed in our previous work
\cite{yerokhin:05:se} for the case of the point nucleus. This method can
be immediately extended to a general (local and spherically-symmetrical)
potential, provided that one can calculate the Green function of the Dirac
equation with this potential. (Beside the full Green
function, the one-potential Green function is also needed in actual
calculations.) In the present work, we develop an
efficient scheme of computation of the Dirac Green function for a general
potential, which is described in Appendix~\ref{app:1} for the full Green
function and Appendix~\ref{app:2} for the one-potential Green function.

The main advantage of the method reported in Ref.~\cite{yerokhin:05:se} is a
fast convergence of the partial-wave expansion of the matrix element (\ref{se1}).
In the present work, we calculate the difference between the point-nucleus and
extended-nucleus matrix elements. For this difference, the partial-wave expansion
converges even faster (especially, in the low-$Z$ region) than for the
self-energy correction. Because of this, we were
able to significantly improve numerical accuracy as compared to
results previously reported in the literature.

Numerical results for the NSE correction to the energy shift
are usually parameterized in the same
way as the one-loop self-energy itself,
\begin{align} \label{7}
\Delta E_{\rm NSE} = \frac{\alpha}{\pi}\, \frac{(\Za)^4}{n^3}\, F_{\rm NSE}(\Za,R)\,.
\end{align}
Comparison of the present results with those by Mohr and Soff
\cite{mohr:93:prl} for the homogeneously charged sphere model is given in
Table~\ref{tab:sefns:compar}. Numerical results obtained in the present work with the Fermi
model of the nuclear charge distribution are summarized in
Table~\ref{tab:sefns}.

The leading dependence of the NSE correction on $R$ and $\Za$ can be
conveniently factorized out in terms of the first-order
NS contribution $\Delta E_{\rm N}$,
\begin{align} \label{8}
\Delta E_{\rm NSE}(njl) = \Delta E_{\rm N}(n\nicefrac12 l)\,\frac{\alpha}{\pi}\,
    {G}_{\rm NSE}(njl)\,,
\end{align}
where $\Delta E_{\rm N}$ is given by Eqs.~(\ref{5}) and (\ref{5a}).
An important feature of this parametrization
\cite{milstein:02:prl,milstein:04}
is that it involves the {\em full} NS correction, rather than only the
leading term of its $\Za$ expansion. With such choice of normalization,
${G}_{\rm NSE}$ is a slowly-varying function of $Z$
and its dependence on $R$ is more tractable.
Note that for the $np_{3/2}$ reference state, Eq.~(\ref{8}) has 
$\Delta E_{\rm N}(np_{1/2})$ as a prefactor, which
was suggested in Ref.~\cite{milstein:04}.
The $\Za$ expansion of the function ${G}_{\rm NSE}$ has the form
\begin{align} \label{8b}
{G}_{\rm NSE}(ns)  = &\, (\Za)\,a_{10}+ (\Za)^2\,\biggl[a_{\rm log}\,\ln \frac{b}{R_{\rm sph}}
 \nonumber \\ &
    + a_{22}\,\ln^2 (\Za)^{-2} + O[\ln (\Za)]\biggr]\, , \\
{G}_{\rm NSE}(np_j) = &\,
    a_{01} \ln (\Za)^{-2} + a_{00}
 \nonumber \\ &
 + (\Za)\,a_{10}+ (\Za)^2\,\biggl[ a_{\rm log}\,\delta_{j,\nicefrac12}\,\ln
   \frac{b}{R_{\rm sph}}
 \nonumber \\ &
 +   a_{21}\,\delta_{j,\nicefrac12}\,\ln (\Za)^{-2}+ O(1)\biggr]\, ,
 \label{8c}
\end{align}
where $b = \exp[1/(2\gamma)-C-5/6]$, $\gamma = \sqrt{1-(\Za)^2}$,
and $C$ is the Euler constant. Known results for the coefficients of the
expansion are listed in Table~\ref{tab:Za}. We note that the logarithmic $a_{22}$ and $a_{21}$ terms
have not yet appeared in the literature.
It was, however, pointed out by Pachucki \cite{pachucki:priv} that
such terms are present and that the coefficients for the
leading logarithms ($\ln^2(\Za)$ for $ns$ states and $\ln(\Za)$ for $np_{1/2}$
states) are the same as for the self-energy 
correction to the hyperfine splitting.
Values of $a_{00}(np_{1/2})$ for $n>2$ can be found in
Ref.~\cite{jentschura:03:jpa}.

\begin{figure*}[thb]
\centerline{\includegraphics[width=\textwidth]{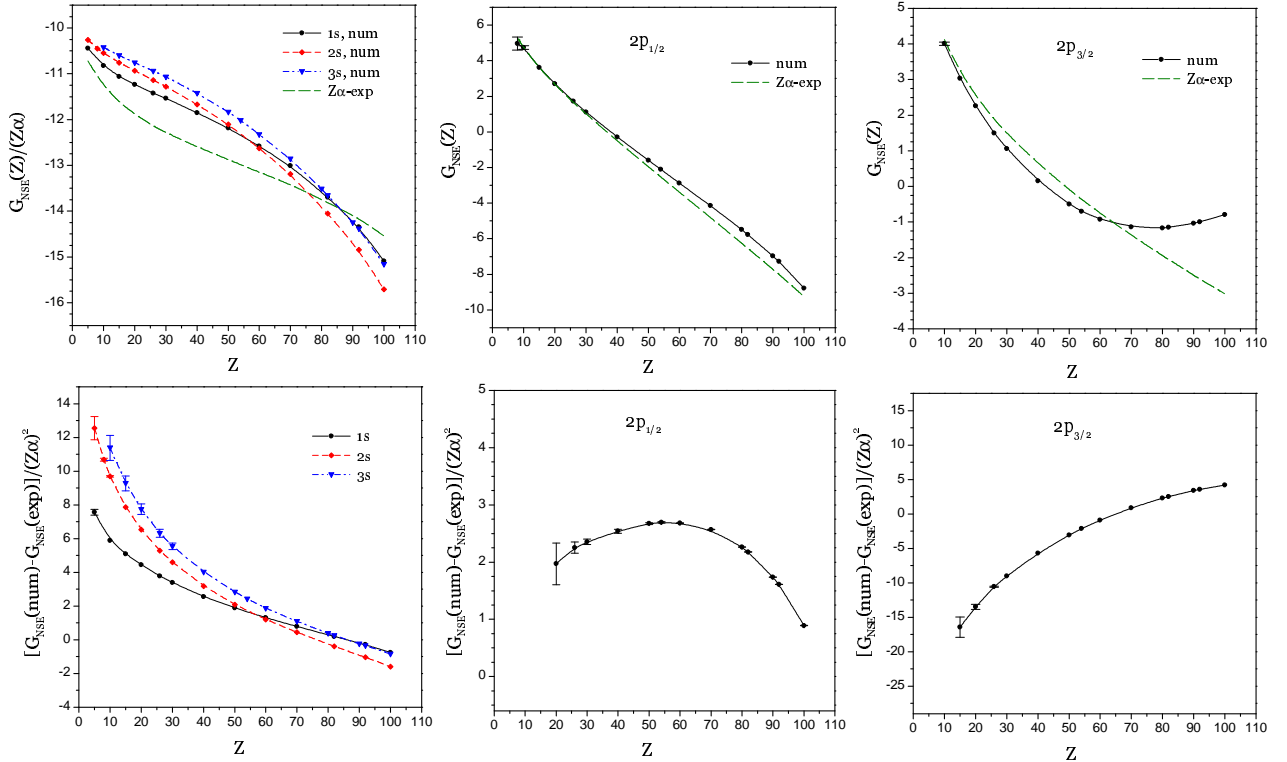}}
\caption{(Color online) Nuclear-size self-energy
correction, in terms of $G_{\rm NSE}$ defined by
  Eq.~(\ref{8}), as a function of the nuclear charge number $Z$.
The upper graphs depict the ratio $G_{\rm NSE}(Z)/(\Za)$ for the $ns$ states and
the function $G_{\rm NSE}(Z)$ for the $np_j$ states, in comparison with the $\Za$-expansion
results. The lower graphs show the difference between the all-order and
$\Za$-expansion results for the function $G_{\rm NSE}$ divided by $(\Za)^2$. 
 \label{fig:sefnsZ} }
\end{figure*}

Comparison of the present numerical data for the function $G_{\rm NSE}$
with the $\Za$-expansion results is given in three upper graphs of
Fig.~\ref{fig:sefnsZ}. Note that for $ns$ states, the ratio
$G_{\rm NSE}(\Za)/(\Za)$ is plotted. The lower graphs in Fig.~\ref{fig:sefnsZ}
depict the higher-order remainder (i.e., the contribution beyond the known
terms of the $\Za$ expansion). For $ns$ states, the remainder does not
approach a finite limit as $Z\to 0$ because it contains $\ln(\Za)$, as can be
seen from Eqs.~(\ref{8b}) and (\ref{8c}). 

\begin{figure*}[thb]
\centerline{\includegraphics[width=\textwidth]{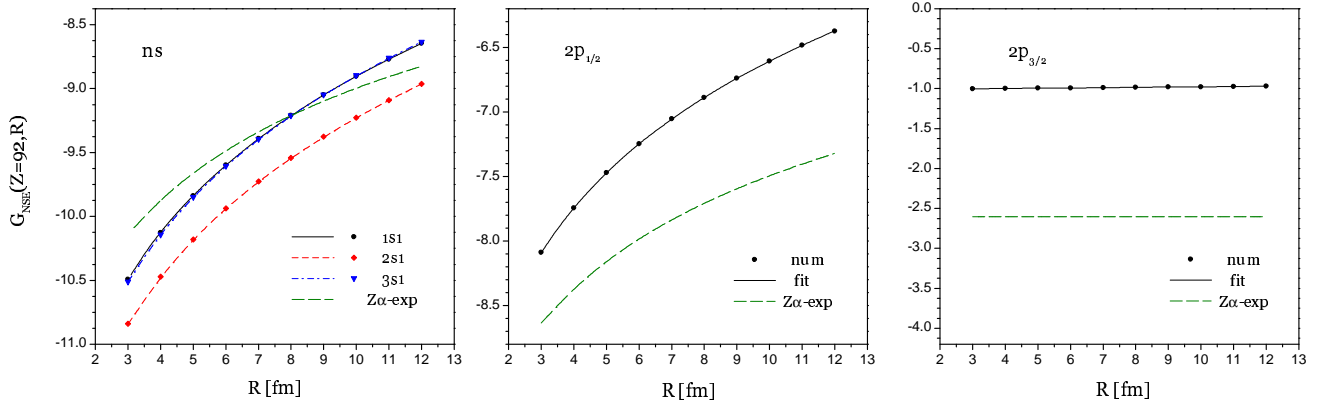}}
\caption{(Color online) Nuclear-size self-energy correction, in terms of $G_{\rm NSE}(\Za,R)$ defined by
  Eq.~(\ref{8}), as a function of the rms nuclear charge radius $R$, for $Z=92$.
 \label{fig:sefnsR} }
\end{figure*}

In Fig.~\ref{fig:sefnsR}, the dependence of $G_{\rm NSE}$ on the rms nuclear
charge radius $R$ is studied, with the nuclear charge number fixed by $Z=92$.
We find that the $R$ dependence of our numerical data can be well approximated by
a three-parameter fit that includes $\ln R$, as suggested by the
$\Za$ expansion.
More specifically, the following fitting functions
approximate the numerical data for $Z=92$ in the region $R = 3$--12~fm with
the relative accuracy of better than $2\times 10^{-4}$
(with $R$ expressed in
fermi units),
\begin{align}
&G_{\rm NSE}(R;1s) = -11.8768 + 1.2083\, \ln R + 0.0191\, R \,,\\
&G_{\rm NSE}(R;2s) = -12.2394 + 1.2124\, \ln R + 0.0220\, R  \,,\\
&G_{\rm NSE}(R;3s) = -11.9131 + 1.2143\, \ln R + 0.0218\, R   \,,\\
&G_{\rm NSE}(R;2p_{1/2}) =
                    -9.4115 + 1.1759\, \ln R + 0.0098\, R  \,,\\
&G_{\rm NSE}(R;2p_{3/2}) =
                    -1.0145 + 0.0019\, \ln R + 0.0033\, R\,.
\end{align}

\section{NS correction to vacuum polarization}

The one-loop vacuum-polarization correction to the energy levels
is usually represented as a sum of
two parts, the Uehling and the Wichmann-Kroll
ones \cite{mohr:98}. The Uehling part of the vacuum polarization is given by the
expectation value of the potential
\begin{align}\label{vp1}
   U_{\rm Ueh}(r) &\ = -\frac{2\alpha^2 Z}{3 m r}\,
    \int_0^{\infty} dr\pr r\pr \rho(r\pr)\,
  \nonumber \\ & \times
       \left[ K_0(2m|r-r\pr|)-K_0(2m|r+r\pr|)\right]\,,
\end{align}
where
\begin{equation}\label{vp2}
 K_0(x) =
      \int_1^{\infty}dt\, e^{-xt}
         \left(\frac1{t^3}+\frac1{2t^5}\right)\,
     \sqrt{t^2-1}\,,
\end{equation}
and the nuclear-charge density $\rho$ is normalized by the condition $\int
d\bm{r}\,\rho(r) = 1$. The energy shift due to the Wichmann-Kroll part
of the vacuum polarization can be written as \cite{soff:88:vp,manakov:89:zhetp}
\begin{align} \label{vp3}
\Delta E_{\rm WK}  = &\,\frac{2\alpha}{\pi}\,{\rm
  Re}\,\sum_{\kappa}|\kappa|\, \int_0^{\infty}d\omega\,
  \int_0^{\infty}r^2dr\, \left( g_a^2+f_a^2\right)\,
 \nonumber \\ & \times
   \int_r^{\infty}dr^{\prime} r^{\prime}\, \left(1-\frac{r^{\prime}}{r}
   \right)\,  {\rm Tr}\, G_{\kappa}^{(2+)}(i\omega,r^{\prime},r^{\prime})\,,
\end{align}
where $g_a$ and $f_a$ are the upper and the lower radial components of the
reference-state wave function, $G_{\kappa}^{(2+)}$ is the radial Dirac
Green function that contains two and more interactions with the binding field,
\begin{align}\label{vp4}
G^{(2+)}_{\kappa}(\omega,x,y) = \int_0^{\infty}&\, dz\, z^2 \,
G^{(0)}_{\kappa}(\omega,x,z)\,
  V(z)\,
  \nonumber \\ \times &\
  \left[G_{\kappa}(\omega,z,y)-G^{(0)}_{\kappa}(\omega,z,y)\right]\,,
\end{align}
$G_{\kappa}$ is the radial part of the full Dirac Green function,
$G^{(0)}_{\kappa}$ is the free Dirac Green function, and $V(z)$ is the binding
potential. We note that Eq.~(\ref{vp3}) is valid both for the point-nucleus
and the extended-nucleus binding potentials.

Calculations of the
Wichmann-Kroll part of the one-loop vacuum polarization were extensively
discussed in the literature over past decades
\cite{soff:88:vp,manakov:89:zhetp,persson:93:vp,sapirstein:03:vp}.
In the present work, we perform calculations of
the vacuum polarization, evaluating the integrations and the summation over
$\kappa$ in the order specified by Eqs.~(\ref{vp3}) and (\ref{vp4}). 
Comparison of the numerical results obtained in this work for the Wichmann-Kroll correction
with those reported in previous calculations \cite{persson:93:vp,mohr:98} is presented in
Table~\ref{tab:wk:compar}.

The nuclear-size vacuum-polarization
(NVP) correction is defined as the difference between the one-loop vacuum-polarization
corrections evaluated with the point-Coulomb potential and the potential of
the extended-charge nucleus.
The NVP correction can be parameterized in the same way as the one-loop radiative corrections,
\begin{align} \label{vp7}
\Delta E_{\rm NVP} = \frac{\alpha}{\pi}\, \frac{(\Za)^4}{n^3}\, F_{\rm NVP}(\Za,R)\,.
\end{align}
The results of our numerical evaluation of the NVP correction for the $1s$,
$2s$, $3s$, $2p_{1/2}$, and $2p_{3/2}$ states of H-like ions are presented in
Table\ref{tab:vpfns}. The calculation
is performed for the Fermi model of the nuclear charge distribution. It is
interesting to note that for the $2p_{3/2}$ state and high nuclear charges,
the correction coming from the Wichmann-Kroll part of the vacuum
polarization dominates over the Uehling part.

The leading dependence of the NVP correction on $R$ and $\Za$ can be
conveniently factorized out in terms of the first-order
NS contribution $\Delta E_{\rm N}$
\cite{milstein:02:prl,milstein:04},
\begin{align} \label{vp8}
\Delta E_{\rm NVP}(njl) = \Delta E_{\rm N}(n\nicefrac12 l)\,\frac{\alpha}{\pi}\,
    {G}_{\rm NVP}(njl)\,.
\end{align}
Note that similarly to the NSE correction,
for the $np_{3/2}$ reference state, Eq.~(\ref{vp8}) has the
first-order NS correction for the $np_{1/2}$ state as a prefactor.

The $\Za$ expansion of the function ${G}_{\rm NVP}$ is given by
\begin{align} \label{vp8b}
{G}_{\rm NVP}(ns) = &\,
 (\Za)\,a_{10}+ (\Za)^2\,\biggl[\frac{2}{3\gamma}\,\ln^2 \frac{b}{R_{\rm sph}}
 \nonumber \\ &
 + a_{21}\,\ln(\Za)^{-2} + f(\Za,R_{\rm sph}) + O(1)\biggr] \,,
\\
{G}_{\rm NVP}(np_j) = &\,
    a_{00}
 + (\Za)\,a_{10}
+ (\Za)^2\,\biggl[\frac{2}{3\gamma}\,\delta_{j,\nicefrac12}\,\ln^2 \frac{b}{R_{\rm sph}}
 \nonumber \\ &
 +  \delta_{j,\nicefrac12}\,f(\Za,R_{\rm sph}) + O(1)\biggr] \,,
\end{align}
where $b = \exp[1/(2\gamma)-C-5/6]$, $\gamma = \sqrt{1-(\Za)^2}$,
and $C$ is the Euler constant. The leading term of the expansion for the $s$
states was calculated long ago \cite{friar:79,hylton:85}. All other coefficients except
$a_{21}$ were derived in Refs.~\cite{milstein:03:fns,milstein:04}. The
logarithmic $a_{21}$ term was pointed out by Pachucki \cite{pachucki:priv};
the value of the coefficient is the same as for the vacuum-polarization
correction to the hyperfine splitting.
The results for the expansion coefficients are
\begin{align} \label{vp9}
&a_{00}(np_{1/2}) = a_{00}(np_{3/2}) = -\nicefrac{8}{45}\,,  \\
&a_{10}(ns) = \nicefrac{3\pi}{4}\,,\ \
a_{10}(np_{1/2}) = \nicefrac{23\pi}{72}\,,\ \
a_{10}(np_{3/2}) = \nicefrac{5\pi}{72}\,,\\
&a_{21}(ns) = \nicefrac{4}{15}\,,
\end{align}
and \cite{milstein:04}
\begin{align} \label{vp10}
f(\Za,R_{\rm sph}) = &\ \frac1{3(\Za)^2}\biggl[
-2\ln R_{\rm sph} -\frac53 +\pi\tan(\pi\gamma)
 \nonumber \\ &
+ \frac{2}{3+2\gamma} +
2\psi(-1-2\gamma)
\nonumber \\ &
-\frac{\pi^{3/2}(3+2\gamma)\Gamma(\gamma+1)}{40\sin(2\pi\gamma)(\gamma-1)
\Gamma(-1-2\gamma)\Gamma(\gamma+3/2)}\,
\nonumber \\ & \times
\left( 2R_{\rm
  sph}\right)^{2(1-\gamma)}
\biggr]\,.
\end{align}
The function $f(\Za,R_{\rm sph})$ has a finite limit as $\Za\to 0$, which is
\begin{align} \label{vp10a}
f(0,R_{\rm sph}) =&\, \frac13\, \ln^2R_{\rm sph}
 + \left(-\frac45 +\frac23\,C \right)\,\ln R_{\rm sph}
\nonumber \\ &
 + \frac13 \left( \frac{833}{255}-\frac{12}{5}\,C+ C^2-\frac{\pi^2}{12}\right)
\,.
\end{align}

\begin{figure*}[thb]
\centerline{\includegraphics[width=\textwidth]{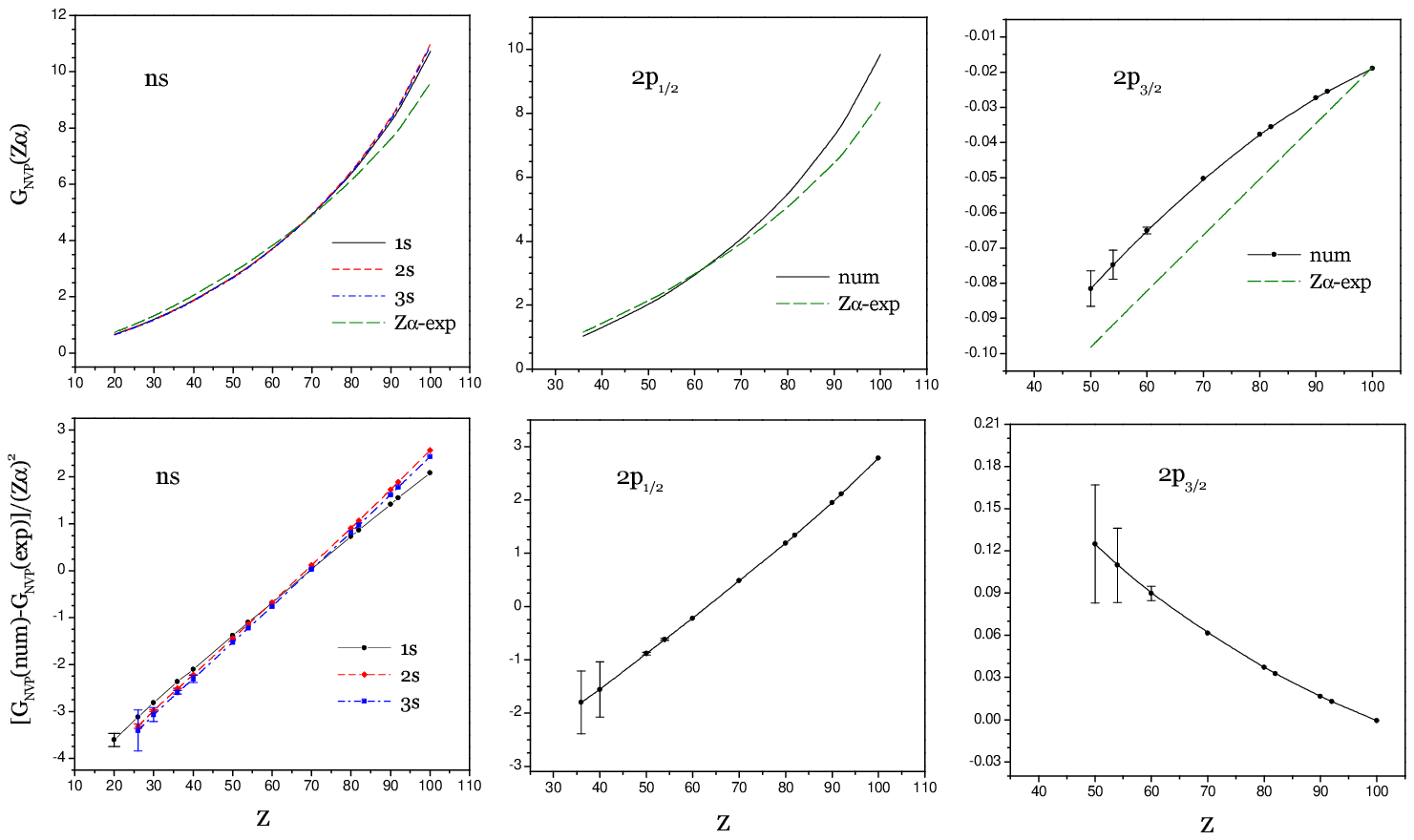}}
\caption{(Color online) Nuclear-size vacuum-polarization  correction, in terms of $G_{\rm NVP}(\Za,R)$ defined by
  Eq.~(\ref{vp7}), as a function of the nuclear charge number $Z$.
 \label{fig:vpfnsZ} }
\end{figure*}

In Fig.~\ref{fig:vpfnsZ}, we compare the present numerical data for the
function $G_{\rm NVP}$ with the $\Za$-expansion results summarized above. We
observe good agreement in all cases; the higher-order remainder function
exhibits a nearly linear dependence on the nuclear charge number.

\begin{figure*}[thb]
\centerline{\includegraphics[width=\textwidth]{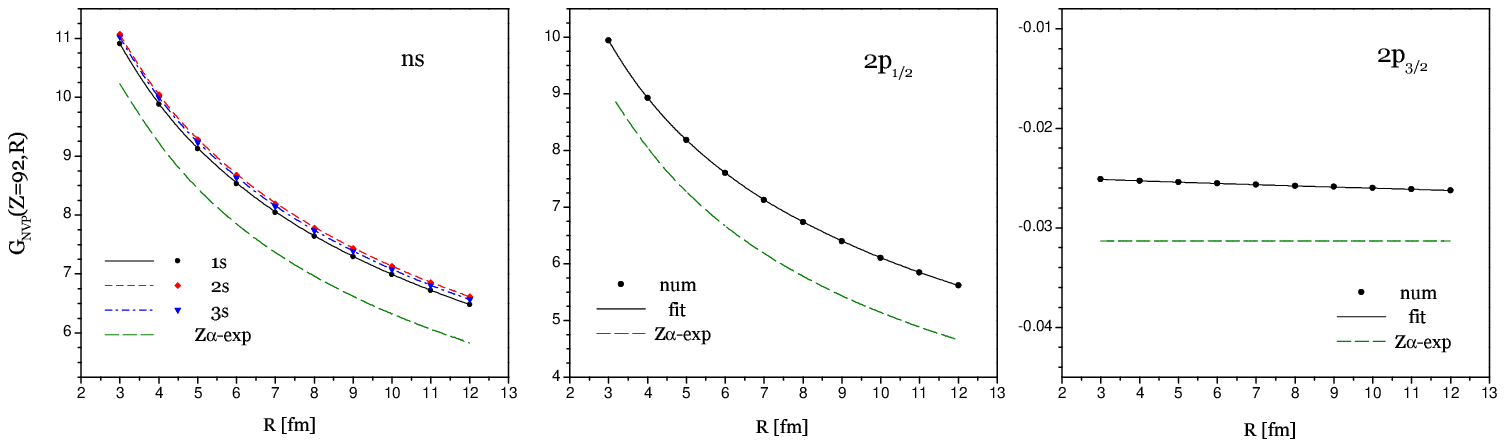}}
\caption{(Color online) Nuclear-size vacuum-polarization correction, in terms of $G_{\rm NVP}(\Za,R)$ defined by
  Eq.~(\ref{vp7}), as a function of the rms nuclear charge radius $R$,  for $Z=92$.
 \label{fig:vpfnsR} }
\end{figure*}

In Fig.~\ref{fig:vpfnsR}, we study the  dependence of the function
$G_{\rm NVP}$ on the rms nuclear
charge radius $R$, with the nuclear charge number fixed by $Z=92$. Similarly
to the NSE correction, we
find that the $R$ dependence of our numerical data can be well approximated by
a three-parameter fit, whose form is suggested by the $\Za$ expansion.
More specifically, the following fitting functions
approximate the numerical data for $Z=92$ in the region $R = 3$--12~fm with
the relative accuracy of better than $2\times 10^{-4}$
(with $R$ expressed in fermi units),
\begin{align}
&G_{\rm NVP}(R;1s) = 15.3607 + 0.3459\, \ln^2 R -4.4325\, \ln R \,,\\
&G_{\rm NVP}(R;2s) = 15.5292 + 0.3397\, \ln^2 R -4.4307\, \ln R \,,\\
&G_{\rm NVP}(R;3s) = 15.4820 + 0.3396\, \ln^2 R -4.4321\, \ln R \,,\\
&G_{\rm NVP}(R;2p_{1/2}) =
                   14.3668 + 0.3673\, \ln^2 R -4.4346\, \ln R \,,\\
&G_{\rm NVP}(R;2p_{3/2}) =
                   -0.02474 - 0.000134\, R + 0.000001\,R^2\,.
\end{align}

\section{Results for hydrogen}

In this section, we obtain all-order (in $\Za$)
results for the radiative nuclear size effect to the ground-state Lamb shift in
hydrogen. This task is complicated by the fact that 
we are not able to perform calculations of the self-energy and 
Wichmann-Kroll parts of the nuclear size effect directly for $Z=1$. 
In the absence of a direct calculation, we perform extrapolation of the 
numerical data obtained for higher values of $Z$ to $Z=1$.

We start with the self energy. The data for the function $G_{\rm
  NSE}(Z,R)$ plotted in Fig.~\ref{fig:sefnsZ} is not well suited for
extrapolation since individual points correspond to different values of the
rms radius. Because of this, we repeat our calculations for different nuclear
charges and the rms radius fixed by $R = r_p$, where $r_p =
0.8768(69)$~fm is the CODATA value of the proton charge radius \cite{mohr:08:rmp}. 
We also account for the fact that the Fermi model of the
nuclear charge distribution is not completely adequate for small rms radii;
the Gaussian model is employed instead, with $\rho(r) = \rho_0\,\exp(-\Lambda r^2)$. The
extrapolation is performed for the higher-order remainder function, 
\begin{equation}
{\cal G}^{\rm h.o.}_{\rm  NSE} = 
[G_{\rm
    NSE}({\rm num})-G_{\rm NSE}({\rm ana})]/(\Za)\,, 
\end{equation}
where $G_{\rm NSE}({\rm num})$ and $G_{\rm NSE}({\rm ana})$ denote the 
numerical and analytical [Eq.~(\ref{8b})]  values of the $G_{\rm  NSE}$ function. 
In our extrapolation, we used 20 points with the nuclear charges in the
interval $Z=5-30$ and
the same extrapolation procedure as in Ref.~\cite{yerokhin:05:hfs}. Our result
for hydrogen is ${\cal G}^{\rm h.o.}_{\rm  NSE}(Z=1) = 0.075(25)$. 

The Uehling part of the NVP correction is calculated directly, with
the result (for the Gaussian model) $G_{\rm NVP, Ue}(Z=1) = 2.5835\,\alpha$. 
The Wichmann-Kroll part is a small correction for hydrogen,
its leading contribution to $G_{\rm NVP}$ being a constant term of 
order $(\Za)^2$. Similarly to the NSE correction, we obtain the result for
hydrogen by extrapolation. The data to be extrapolated is obtained by repeating
our calculations for $Z= 20-75$, with the rms radius fixed by $R = r_p$
and the nuclear charge distribution given by the
Gaussian model. The
extrapolation is performed for the ratio $G_{\rm NVP,WK}/(\Za)^2$. The
result for hydrogen is $G_{\rm NVP, WK}(Z=1) = -9.8(9)\,\alpha^2$.

Summarising our calculations of the radiative nuclear size effect to the
$1s$ Lamb shift in hydrogen, we express the results in the same form as in
Ref.~\cite{mohr:08:rmp}, 
\begin{align}  \label{x1}
&\Delta E_{\rm NSE} = \alpha^2 Z\,{\cal E}_{\rm N}\,\bigl[ -3.1294(80)\bigr]\,,
\\
\label{x2}
&\Delta E_{\rm NVP} = \alpha^2Z\,{\cal E}_{\rm N}\,\bigl[  0.8228 -0.0228(23)\bigr]\,,
\end{align}
where ${\cal E}_{\rm N} = 2/3\, (\Za)^4\,R^2$ and the first and the second
terms in the brackets in Eq.~(\ref{x2}) correspond to the Uehling and
Wichmann-Kroll parts. In order to estimate the model dependence of our
results, we evaluated the Uehling part also within the exponential model,
with $\rho(r) = \rho_0\,\exp(-\Lambda\,r)$, and found a 0.2\% deviation from the 
Gaussian result. 

The numerical constant terms in Eqs.~(\ref{x1}) and (\ref{x2}) can be compared
with the $\Za$-expansion results. For the self-energy, the leading-order term of 
the $\Za$ expansion is $4\ln2-\nicefrac{23}{4} = -2.977$, whereas all terms 
in Eq.~(\ref{8b}) yield the coefficient of $-3.153$. For the vacuum polarization,
the leading-order term is $\nicefrac34$, whereas all terms 
in Eq.~(\ref{vp8b}) yield $0.827$. We conclude that the higher-order
corrections increase the leading-order result for 
the radiative nuclear size effect in hydrogen by 4.4\%.

\section*{Conclusion}
\label{sec:concl}

In the present investigation, we evaluate the nuclear size correction
to the Lamb shift of the $1s$, $2s$, $3s$, $2p_{1/2}$, and $2p_{3/2}$ states
of hydrogen-like atoms. The treatment is complete at the one-loop level, i.e.,
it includes the leading-order effect as well as the one-loop radiative
corrections. The total nuclear size correction to the energy level
is represented, for the $ns$ and $np_{1/2}$ states, as
\begin{align}
\Delta E_{\rm NS} = \Delta E_{\rm N} \, \biggl[ 1+ \frac{\alpha}{\pi}\,\left(G_{\rm NSE}+
  G_{\rm NVP}\right) \biggr]\,,
\end{align}
and, for the $np_{3/2}$ states, as
\begin{align}
\Delta E_{\rm NS}(np_{3/2}) = &\,\Delta E_{\rm N}(np_{3/2})
\nonumber \\ &
+
 \Delta E_{\rm N}(np_{1/2}) \, \frac{\alpha}{\pi}\,\left( G_{\rm NSE}+
  G_{\rm NVP} \right)\,,
\end{align}
where $\Delta E_{\rm N}$ is the nuclear size correction to the Dirac energy.
The all-order numerical values obtained for the self-energy and
vacuum-polarization functions $G_{\rm NSE}$ and $G_{\rm NVP}$ were compared
with results of the $\Za$-expansion calculations. Inclusion of the logarithmic
term of the relative order $(\Za)^2 \ln^2 (\Za)^{-2}$ for $ns$ states
was necessary in order to achieve agreement between different calculations.
Extrapolation of the all-order data obtained for hydrogen-like ions to
$Z=1$ provides an all-order result for the radiative nuclear size effect on
the ground-state Lamb shift in hyrogen. The higher-order corrections are shown
to increase the leading-order result by 4.4\%.


\section*{Acknowledgments}

I wish to thank Krzysztof Pachucki for valuable comments and advices.
Computations reported in this work were performed on the
computer cluster of St.~Petersburg State Polytechnical University.

%
%
\begin{table*}[htb]
\caption{Nuclear-size correction to the Dirac energies of H-like ions, in terms of
the function  $G_{\rm N}(\Za,R)$ defined by Eq.~(\ref{5}). Fermi model of the nuclear
  charge distribution is used.
 \label{tab:fns}}
\begin{ruledtabular}
\begin{tabular}{cc.....}
 $Z$ & $R$ [fm] &\multicolumn{1}{c}{$1s$} &
                     \multicolumn{1}{c}{$2s$} &
                                   \multicolumn{1}{c}{$3s$} &
                                          \multicolumn{1}{c}{$2p_{1/2}$}  \\
\hline\\[-7pt]
   5 &   2.4059 &   1.00x0\,46 &   1.00x0\,71 &   1.00x0\,23 &   1.00x1\,73\\
   8 &   2.7013 &   1.00x3\,91 &   1.00x4\,55 &   1.00x3\,33 &   1.00x6\,89\\
  10 &   3.0053 &   1.00x6\,57 &   1.00x7\,58 &   1.00x5\,67 &   1.01x1\,17\\
  15 &   3.1888 &   1.01x5\,66 &   1.01x7\,93 &   1.01x3\,60 &   1.02x5\,66\\
  20 &   3.4764 &   1.02x8\,67 &   1.03x2\,74 &   1.02x4\,91 &   1.04x6\,42\\
  26 &   3.7371 &   1.04x9\,77 &   1.05x6\,75 &   1.04x3\,18 &   1.08x0\,18\\
  30 &   3.9286 &   1.06x7\,32 &   1.07x6\,73 &   1.05x8\,28 &   1.10x8\,52\\
  40 &   4.2696 &   1.12x4\,66 &   1.14x2\,02 &   1.10x6\,96 &   1.20x2\,64\\
  50 &   4.6543 &   1.20x3\,59 &   1.23x2\,01 &   1.17x2\,31 &   1.33x7\,09\\
  60 &   4.9118 &   1.30x8\,62 &   1.35x1\,81 &   1.25x6\,25 &   1.52x4\,64\\
  70 &   5.3115 &   1.44x5\,02 &   1.50x7\,15 &   1.35x9\,74 &   1.78x4\,78\\
  82 &   5.5010 &   1.66x2\,15 &   1.75x2\,74 &   1.51x1\,54 &   2.23x6\,31\\
  92 &   5.8569 &   1.89x6\,75 &   2.01x3\,31 &   1.65x5\,09 &   2.78x5\,73\\
 100 &   5.8570 &   2.12x8\,53 &   2.26x3\,06 &   1.77x4\,54 &   3.39x3\,88\\
\end{tabular}
\end{ruledtabular}
\end{table*}

%
%
\begin{table*}[htb]
\caption{Different calculations of the
nuclear-size self-energy correction to the energy levels of H-like ions, in terms of
the function  $F_{\rm NSE}(\Za,R)$ defined by Eq.~(\ref{7}), for the homogeneously charged
  sphere model.
 \label{tab:sefns:compar}}
\begin{ruledtabular}
\begin{tabular}{cc...c}
 $Z$ & $R$ [fm]& \multicolumn{1}{c}{$1s$} & \multicolumn{1}{c}{$2s$} &
  \multicolumn{1}{c}{$2p_{1/2}$} & Ref. \\
\hline\\[-7pt]
26 & 3.730 &  -0.000\,17x2\,122(4) & -0.000\,1x73\,29(1)&    0.000\,0x00\,972(2) &\\
   &       &  -0.000\,17x2(1)\,    & -0.000\,1x8(1)  \, &   -0.000\,0x0(1)  &  \cite{mohr:93:prl}\\
54 & 4.826 &  -0.001\,27x4\,870(2) & -0.001\,4x61\,69(1)&   -0.000\,0x20\,391(2) &\\
   &       &  -0.001\,27x5(1)  \,  & -0.001\,4x62(1)  \,&   -0.000\,0x21(1) &  \cite{mohr:93:prl} \\
92 & 5.863 &  -0.018\,49x1\,8(8)   & -0.029\,0x89\,3(9) &   -0.002\,4x82\,76(8) &\\
   &       &  -0.018\,49x2(1)  \,  & -0.029\,0x90(2) \, &   -0.002\,4x83(1) &  \cite{mohr:93:prl} \\
\end{tabular}
\end{ruledtabular}
\end{table*}

%
%
\begin{table*}[htb]
\caption{Nuclear-size self-energy correction to the energy levels of H-like ions, in terms of
the function $F_{\rm NSE}(\Za,R)$ defined by Eq.~(\ref{7}). Fermi model of the nuclear
  charge distribution is used. The nuclear charge rms radii used are listed in Table~\ref{tab:fns}.
 \label{tab:sefns}}
\begin{ruledtabular}
\begin{tabular}{c.....}
 $Z$ & \multicolumn{1}{c}{$1s$} &
                     \multicolumn{1}{c}{$2s$} &
                                   \multicolumn{1}{c}{$3s$} &
                                          \multicolumn{1}{c}{$2p_{1/2}$} &
                                                  \multicolumn{1}{c}{$2p_{3/2}$} \\
\hline\\[-7pt]
  5 & -0.000\,0x09\,962(6) & -0.000\,0x09\,81(2)  &       \, x    \,     &     \,  x  \,        &     \,  x \,         \\
  8 & -0.000\,0x20\,969(2) & -0.000\,0x20\,53(1)  &       \, x    \,     & 0.000\,0x00\,107(8)  &     \,  x \,         \\
 10 & -0.000\,0x33\,242(2) & -0.000\,0x32\,54(1)  & -0.000\,0x32\,12(3)  & 0.000\,0x00\,201(4)  &0.000\,0x00\,170(2)  \\
 15 & -0.000\,0x60\,128(2) & -0.000\,0x59\,10(1)  & -0.000\,0x58\,21(4)  & 0.000\,0x00\,412(2)  &0.000\,0x00\,345(2)  \\
 20 & -0.000\,1x02\,940(2) & -0.000\,1x02\,06(2)  & -0.000\,1x00\,38(2)  & 0.000\,0x00\,699(2)  &0.000\,0x00\,587(2)  \\
 26 & -0.000\,1x72\,537(2) & -0.000\,1x73\,71(3)  & -0.000\,1x70\,72(3)  & 0.000\,0x00\,976(2)  &0.000\,0x00\,852(2)  \\
 30 & -0.000\,2x38\,936(2) & -0.000\,2x43\,68(3)  & -0.000\,2x39\,45(3)  & 0.000\,0x01\,021(2)  &0.000\,0x00\,969(2)  \\
 40 & -0.000\,4x81\,097(2) & -0.000\,5x11\,07(2)  & -0.000\,5x02\,4(1)   &-0.000\,0x00\,726(2) & 0.000\,0x00\,396(2)  \\
 50 & -0.000\,9x60\,607(2) & -0.001\,0x75\,22(2)  & -0.001\,0x57\,8(1)   &-0.000\,0x10\,544(2) &-0.000\,0x03\,299(2) \\
 60 & -0.001\,8x30\,75(1)  & -0.002\,1x83\,28(2)  & -0.002\,1x50\,3(4)   &-0.000\,0x46\,025(4) &-0.000\,0x14\,862(2) \\
 70 & -0.003\,7x30\,19(4)  & -0.004\,7x91\,85(6)  & -0.004\,7x22\,0(2)   &-0.000\,1x70\,450(6) &-0.000\,0x47\,078(2) \\
 82 & -0.008\,4x04\,0(2)   & -0.011\,9x77\,2(2)   & -0.011\,7x96\,6(4)   &-0.000\,7x03\,91(2)  &-0.000\,1x40\,462(2) \\
 92 & -0.018\,4x26\,6(7)   & -0.028\,9x86\,6(8)   & -0.028\,4x74\,0(6)   &-0.002\,4x73\,96(6)  &-0.000\,3x37\,183(2) \\
100 & -0.034\,0x60(2)      & -0.058\,4x44(3)      & -0.057\,1x72(1)\,    &-0.006\,6x96\,3(4)   &-0.000\,6x02\,404(2)
\end{tabular}
\end{ruledtabular}
\end{table*}

%
%
\begin{table}
\caption{Coefficients of the $\Za$ expansion of the NSE correction in
  Eq.~(\ref{8b}).
 \label{tab:Za}}
\begin{ruledtabular}
\begin{tabular}{lccc}
Term & State & Value & Ref. \\ \hline\\[-7pt]
 $a_{01}$  & $np_j$  & $\nicefrac89$  & \cite{milstein:03:fns,jentschura:03:jpa} \\
 $a_{00}$  & $2p_{1/2}$ & 0.808\,879\,967(1) & \cite{milstein:03:fns,jentschura:03:jpa} \\
           & $np_{3/2}$ & $a_{00}(np_{1/2})-1$  & \cite{milstein:03:fns,jentschura:03:jpa} \\
 $a_{10}$  & $ns$      & $\pi \left(  - \nicefrac{23}{4}+ 4\ln 2\right)$ & \cite{pachucki:93:density,eides:97:pra,milstein:02:prl}\\
           & $np_{1/2}$ & $\pi \left( \nicefrac{379}{432}- \nicefrac{16}{3}\ln 2
                               \right)$ & \cite{milstein:04}\\
           & $np_{3/2}$ & $\pi \left( \nicefrac{559}{432}- 4 \ln 2 \right)$ & \cite{milstein:04}\\
 $a_{\log}$& $ns$, $np_{1/2}$ & $\nicefrac{\pi^2}{6}-\nicefrac{15}{4}$ & \cite{milstein:02:prl,milstein:03:fns}\\
 $a_{22}$  & $ns$          & $-\nicefrac23$ & \cite{pachucki:priv}\\
 $a_{21}$  & $np_{1/2}$     & $-2(n^2-1)/n^2$ & \cite{pachucki:priv,jentschura:10}
\end{tabular}
\end{ruledtabular}
\end{table}

%
%

%
%

%
%
\begin{table*}[htb]
\caption{Different calculations of the Wichmann-Kroll vacuum-polarization correction
[Eq.~(\ref{vp3})] to the energy levels of H-like ions, in units
 $\Delta E/[(\alpha/\pi)\,(\Za)^4/n^3]$.
 \label{tab:wk:compar}}
\begin{ruledtabular}
\begin{tabular}{ccl....c}
 $Z$ & $R$ [fm]& Model & \multicolumn{1}{c}{$1s$} & \multicolumn{1}{c}{$2s$} &
  \multicolumn{1}{c}{$2p_{1/2}$} & \multicolumn{1}{c}{$2p_{3/2}$} & Ref. \\
\hline\\[-7pt]
36 & 4.230 & sphere & 0.002\,7x41\,99(6) & 0.002\,8x33\,98(3) & 0.000\,0x83\,36(4) & 0.000\,0x27\,95(4)&\\
   &       & sphere & 0.002\,7x41\,8     & 0.002\,8x33\,7     & 0.000\,0x83\,4     & 0.000\,0x28\,0 &\cite{persson:93:vp}\\
   &       & shell  & 0.002\,7x  \,      & 0.002\,8x  \,      & 0.000\,1x  \,      & 0.000\,0x  \,&\cite{mohr:98}\\
54 & 4.826 & sphere & 0.005\,9x21\,2(2)  & 0.006\,4x23\,4(2)  & 0.000\,4x54\,79(2) & 0.000\,1x14\,18(5)&\\
   &       & sphere & 0.005\,9x21\,1     & 0.006\,4x23\,1     & 0.000\,4x54\,8     & 0.000\,1x14\,2 &\cite{persson:93:vp}\\
   &       & shell  & 0.005\,9x  \,      & 0.006\,4x  \,      & 0.000\,4x  \,      & 0.000\,1x \,&\cite{mohr:98}\\
92 & 5.860 & sphere & 0.020\,6x79\,2(5)  & 0.027\,2x52\,0(8)  & 0.006\,8x24\,1(4)  & 0.000\,7x49\,49(6)&\\
   &       & sphere & 0.020\,6x78\,9     & 0.027\,2x51\,5     & 0.006\,8x24\,0     & 0.000\,7x49\,4 &\cite{persson:93:vp}\\
   &       & shell  & 0.020\,6x  \,      & 0.027\,2x  \,      & 0.006\,8x  \,      & 0.000\,7x \,&\cite{mohr:98}\\
\end{tabular}
\end{ruledtabular}
\end{table*}

%
%
\begin{table*}[htb]
\caption{Nuclear-size vacuum-polarization
correction to the energy levels of H-like ions, in terms of
the function $F_{\rm NVP}(\Za,R)$ defined by Eq.~(\ref{vp7}). Fermi model of the nuclear
charge distribution is used. The nuclear charge rms radii used are listed in Table~\ref{tab:fns}.
For a given $Z$, the upper line (``Ue'') corresponds to the Uehling part and
the lower line (``WK''), to the Wichmann-Kroll part.
 \label{tab:vpfns}}
\begin{ruledtabular}
\begin{tabular}{cc.....}
 $Z$ &  & \multicolumn{1}{c}{$1s$} &
                                   \multicolumn{1}{c}{$2s$} &
                                           \multicolumn{1}{c}{$3s$} &
                                                    \multicolumn{1}{c}{$2p_{1/2}$} &
                                                                  \multicolumn{1}{c}{$2p_{3/2}$}
                                                                  \\
\hline\\[-7pt]
 15 & Ue &  0.000\,0x24\,856   &  0.000\,0x24\,968   &  0.000\,0x24\,921   &  0.000\,0x00\,020   & -0.000\,0x00\,016   \\[2pt]
 20 & Ue &  0.000\,0x47\,62    &  0.000\,0x48\,21    &  0.000\,0x48\,12    &  0.000\,0x00\,102   & -0.000\,0x00\,034   \\
    & WK & -0.000\,0x06\,4(2)  & -0.000\,0x06(1)\,   & -0.000\,0x06(3)\,   &       \, x  \,      &       \, x  \,      \\[2pt]
 26 & Ue &  0.000\,0x89\,44    &  0.000\,0x91\,72    &  0.000\,0x91\,60    &  0.000\,0x00\,402   & -0.000\,0x00\,064   \\
    & WK & -0.000\,0x12\,58(1) & -0.000\,0x12\,9(1)  & -0.000\,0x13(1)\,   &       \, x  \,      &       \, x  \,      \\[2pt]
 30 & Ue &  0.000\,1x31\,907(2)&  0.000\,1x36\,725(2)&  0.000\,1x36\,601(2)&  0.000\,0x00\,865   & -0.000\,0x00\,092   \\
    & WK & -0.000\,0x18\,83(3) & -0.000\,0x19\,55(9) & -0.000\,0x19\,6(7)  &       \, x  \,      &       \, x  \,      \\[2pt]
 40 & Ue &  0.000\,3x04\,304(4)&  0.000\,3x26\,352(4)&  0.000\,3x26\,510(4)&  0.000\,0x04\,205   & -0.000\,0x00\,188   \\
    & WK & -0.000\,0x43\,43(4) & -0.000\,0x46\,5(1)  & -0.000\,0x46\,6(9)  & -0.000\,0x00\,9(1)  &       \, x  \,      \\[2pt]
 50 & Ue &  0.000\,6x74\,503(2)&  0.000\,7x56\,416(2)&  0.000\,7x58\,099(2)&  0.000\,0x16\,672   & -0.000\,0x00\,342   \\
    & WK & -0.000\,0x92\,38(7) & -0.000\,1x02\,71(7) & -0.000\,1x03\,07(9) & -0.000\,0x03\,37(2) & -0.000\,0x00\,20(2) \\[2pt]
 60 & Ue &  0.001\,4x10\,95(1) &  0.001\,6x73\,01(1) &  0.001\,6x80\,03(1) &  0.000\,0x57\,40    & -0.000\,0x00\,546   \\
    & WK & -0.000\,1x81\,6(2)  & -0.000\,2x11\,8(3)  & -0.000\,2x12\,8(3)  & -0.000\,0x10\,46(2) & -0.000\,0x00\,49(2) \\[2pt]
 70 & Ue &  0.003\,1x00\,32(1) &  0.003\,9x32\,42(2) &  0.003\,9x56\,23(2) &  0.000\,1x97\,79    & -0.000\,0x00\,872   \\
    & WK & -0.000\,3x67\,7(4)  & -0.000\,4x54\,6(5)  & -0.000\,4x56\,8(5)  & -0.000\,0x32\,06(4) & -0.000\,0x01\,201(4)\\[2pt]
 82 & Ue &  0.007\,7x10\,07(4) &  0.010\,7x79\,92(6) &  0.010\,8x63\,65(6) &  0.000\,8x22\,122(5)& -0.000\,0x01\,310   \\
    & WK & -0.000\,8x21\,5(7)  & -0.001\,1x06\,2(9)  & -0.001\,1x11\,4(9)  & -0.000\,1x14\,34(9) & -0.000\,0x03\,021(4)\\[2pt]
 92 & Ue &  0.018\,2x30\,65    &  0.028\,0x56\,439(2)&  0.028\,2x75\,306(4)&  0.002\,9x70\,972   & -0.000\,0x01\,923   \\
    & WK & -0.001\,7x62\,6(8)  & -0.002\,5x87(1) \,  & -0.002\,5x94(1)\,   & -0.000\,3x61\,0(2)  & -0.000\,0x06\,750(3)\\[2pt]
100 & Ue &  0.036\,4x29\,910(6)&  0.061\,1x65\,53(1) &  0.061\,5x49\,11(1) &  0.008\,4x27\,011(2)& -0.000\,0x02\,344   \\
    & WK & -0.003\,2x59(2) \,  & -0.005\,1x84(4)\,   & -0.005\,1x83(4)\,   & -0.000\,9x15\,4(6)  & -0.000\,0x12\,100(6)
\end{tabular}
\end{ruledtabular}
\end{table*}

%
%

%
%

\appendix

\section{Dirac Green function for a general potential}
\label{app:1}

In this section we construct the Green function of the Dirac equation with the
potential $V(r)$ of a general form.
We assume that the potential $V(r)$ differs from
the Coulomb one within a finite (inner) region only, i.e., there is $r_0$ such
that, for $r>r_0$, $V(r) = -Z\alpha/r$, with $Z \geq 0$. For our purposes, it
is sufficient to consider the potential to be regular at the origin,
i.e., $rV(r)\to 0$ as $r\to 0$. In the inner region
$r<r_0$, the combination $rV(r)$ is assumed to be well represented by a piecewise
cubic polynomial calculated on a sufficiently dense grid.

The radial Dirac Green function $G_{\kappa}(E,r_1,r_2)$ is expressed in terms
of the two-component solutions of the radial Dirac equation
regular at the origin $\left(\phi_{\kappa}^{0}\right)$
and the infinity $\left(\phi_{\kappa}^{\infty}\right)$
as follows
\begin{align}\label{gr01}
 G_{\kappa}(E,r_1,r_2) = &\,
 -\phi_{\kappa}^{\infty}(E,r_1)\,\phi_{\kappa}^{0^T}(E,r_2)\,\theta(r_1-r_2)
\nonumber \\ &
 -\phi_{\kappa}^{0}(E,r_1)\,\phi_{\kappa}^{{\infty}^T}(E,r_2)\,\theta(r_2-r_1)\,.
\end{align}
The solutions are normalized by the condition that their Wronskian is unity
(everywhere except for the bound-state energies),
\begin{align}\label{gr02}
 \phi_{\kappa}^{0^T}(E,r) \Dmatrix{0}{-1}{1}{0} \phi_{\kappa}^{\infty}(E,r) = 1\,.
\end{align}

In the present work, we obtain the radial solutions in the inner region
$r<r_0$ by a numerical solution of the Dirac equation on a grid, whereas in
the outer region $r>r_0$, we express them as a combination of the radial
solutions of the Dirac-Coulomb problem. The regular and irregular solutions
of the Dirac equation with the point-nucleus Coulomb potential will be denoted
by $\psi_{\kappa}^{0}$ and  $\psi_{\kappa}^{\infty}$, respectively. They are
known analytically in terms of the Whittaker functions,
see e.g., Ref.~\cite{mohr:98}. (Note the sign difference of the
present definition of the Green function as compared to that of Ref.~\cite{mohr:98}.)
In this work, the
Dirac-Coulomb solutions $\psi_{\kappa}^{0}$ and  $\psi_{\kappa}^{\infty}$
are evaluated by a generalization of the procedure described in Ref.~\cite{yerokhin:99:pra}.

The general calculational scheme is as follows. For a given energy argument
$E$, we calculate and store the solutions $\psi_{\kappa}^{0}$ and
$\psi_{\kappa}^{\infty}$ on a radial grid $\left\{r_i\right\}_{i=1}^N$ and
then obtain the radial Green function for arbitrary radial arguments
by interpolation. Large number of the
mesh points ($N\sim 10^4$) and a careful choice of the grid allow us to
minimize losses of accuracy due to interpolation. In order to avoid
numerical overflow (underflow) when storing the regular and irregular
solutions for large imaginary
energies $E$ and large $\kappa$, all manipulations are performed
with the ``normalized'' solutions in which the approximate large-$r$ and small-$r$
behaviour is pulled out,
\begin{align}\label{gr03}
& \widetilde{\phi}_{\kappa}^{0}(E,r) =
    r^{-|\kappa|}\,e^{-cr}\,\phi_{\kappa}^{0}(E,r)\,, \\
& \widetilde{\phi}_{\kappa}^{\infty}(E,r) =
    r^{|\kappa|}\,e^{cr}\,\phi_{\kappa}^{\infty}(E,r)\,,
\label{gr03a}
\end{align}
where $c = \sqrt{1-E^2}$. Advantages of the normalized solutions are, first,
that they are more suitable for interpolation than the original solutions and,
more importantly, that they can be stored and manipulated within the  standard double
precision arithmetics (in the range of $\kappa$'s relevant for the present
investigation, up to $|\kappa|\sim 30$).

In the inner region $r<r_0$, we calculate the regular solution
$\phi_{\kappa}^{0}$ (or, rather, $\widetilde{\phi}_{\kappa}^{0}$)
by solving the radial Dirac equation on a grid as described in the following,
starting from $r=0$ and up to $r=r_0$. For $r>r_0$, the potential is the
Coulomb one and the regular solution $\phi_{\kappa}^{0}$ is a linear
combination of the regular and irregular Dirac-Coulomb solutions,
\begin{align}\label{gr04}
\phi_{\kappa}^{0} (E,r) = a\, \psi_{\kappa}^{0} (E,r)+b\,
\psi_{\kappa}^{\infty} (E,r)\,,\ \ \ r\geq r_0\,.
\end{align}
The coefficients $a$ and $b$ are defined by the condition that the two
components of $\phi_{\kappa}^{0}$ are continuous at $r=r_0$. So, we determine
the coefficients $a$ and $b$ by matching the numerical and the analytical
solutions at $r=r_0$ and employ the analytical Dirac-Coulomb functions for
calculations for $r>r_0$.

The irregular solution $\phi_{\kappa}^{\infty}$ in the outer region is just
the Dirac Coulomb function,
\begin{align}\label{gr05}
\phi_{\kappa}^{\infty} (E,r) = \psi_{\kappa}^{\infty} (E,r)\,,\ \ \ r\geq r_0\,.
\end{align}
So, we use the analytical representation for $r\geq r_0$. For smaller
$r$, the irregular solution is calculated by solving the Dirac equation on a
grid, downward from $r=r_0$  towards $r=0$. The normalization of the numerical
solution is fixed by requiring continuity at the point
$r = r_0$.

We now turn to the problem of solving the Dirac equation with the potential $V(r)$
on a grid. In this work, we employ the power
series solution method, previously applied to the Dirac equation
by Salvat et al.~\cite{salvat:95:cpc}. For completeness, we give here the
description of the method. First, let us solve the equation
on the interval $[r_a,r_b]$
with given boundary conditions at $r=r_a$.
The situation $r_b<r_a$ is allowed and it is assumed that $r_a>0$. (The
special case of $r_a = 0$ will be considered separately.)
The radial Dirac equation is (with $m=1$)
\begin{align} \label{a1}
G^{\prime}(r) &\ = -\frac{\kappa}{r}\,G(r) + (E-V(r)+1)\,F(r)\,,\\
F^{\prime}(r) &\ =  \frac{\kappa}{r}\,F(r) - (E-V(r)-1)\,G(r)\,,
\end{align}
where
$G(r) = rg(r)$ and $F(r) = rf(r)$ are the upper and lower components of
the radial Dirac solution.
Introducing new
variables $x = (r-r_a)/h$ and $h = r_b-r_a$, the equation is written as
\begin{align} \label{a3}
(xh+r_a)G^{\prime}_x + \kappa h G + U h F - (xh+r_a) h F &\ = 0\,,\\
(xh+r_a)F^{\prime}_x - \kappa h F - U h G - (xh+r_a) h G &\ = 0\,,
\end{align}
where $U = r(V(r)-E)$. On the given interval, $U$ is represented by a cubic
polynomial of $x$, $U = \sum_{k=0}^3
u_k x^k$. The solutions are
represented as power series of the form
\begin{align} \label{a4}
G(x) = \sum_{n = 0}^{n_{\rm max}} a_n x^n\,,\ \ \
F(x) = \sum_{n = 0}^{n_{\rm max}} b_n x^n\,,
\end{align}
with the coefficients $a_0$ and $b_0$ determined by the boundary conditions
$a_0 = G(r_a)$ and $b_0 = F(r_a)$.
The coefficients $a_n$ and $b_n$ are determined by the recurrence relations
(valid for $r_a \neq 0$)
\begin{align} \label{a5}
a_n &\ = -\frac{h}{nr_a}\bigl[(n-1+\kappa)a_{n-1}+(u_0-r_a)b_{n-1}
\nonumber \\ &
  +(u_1-h)b_{n-2}+u_2b_{n-3}+ u_3b_{n-4} \bigr]\,, \\
b_n &\ =  \frac{h}{nr_a}\bigl[(-n+1+\kappa)b_{n-1}+(u_0+r_a)a_{n-1}
\nonumber \\ &
  +(u_1+h)a_{n-2}+u_2a_{n-3}+ u_3a_{n-4} \bigr]\,.
\label{a6a}
\end{align}
The solutions at the end point are given by the sum of the coefficients,
\begin{align} \label{a6}
G(r_b) = \sum_{n = 0}^{n_{\rm max}} a_n \,, \ \ \
F(r_b) = \sum_{n = 0}^{n_{\rm max}} b_n\,.
\end{align}
In the numerical evaluation, the recurrence relations are applied upwards
until either the desired precision or the upper limit for $n$
(typically, $n_{\rm max}=30$) is reached. In the latter case,
the interval is subdivided into two parts and the procedure is
repeated until the desired accuracy is attained. This simple approach allows
one to solve the equation with accuracy close to the machine precision.

Now, we consider the special case of $r_a = 0$. In this case, the solutions
are represented by the power expansion of the form
\begin{align} \label{a7}
G(x) = x^s\,\sum_{n = 0}^{n_{\rm max}} a_n x^n\,,\ \ \
F(x) = x^{s+t}\,\sum_{n = 0}^{n_{\rm max}} b_n x^n\,,
\end{align}
where the parameters $s$ and $t$ are determined from the Dirac equation. For
$\kappa < 0$, we have (for the regular potentials considered here)
$s = |\kappa|$ and $t = 1$. The series start with
\begin{align} \label{a8}
a_0 = 1\,, \ \ \ b_0 = \frac{h+u_1}{1+2|\kappa|}\,.
\end{align}
The recursion relations take the form
\begin{align} \label{a9}
& n a_n  = (u_1-h) b_{n-2}+ u_2 b_{n-3}+ u_3 b_{n-4}\,,\\
& (2|\kappa|+n+1) b_n  = (h+u_1)a_n+ u_2a_{n-1} + u_3a_{n-2}\,.
\end{align}
For $\kappa >0$, one gets $s = \kappa+1$ and $t = -1$. The series start with
\begin{align} \label{a10}
a_0 = \frac{h-u_1}{1+2\kappa}\,, \ \ \ b_0 = 1\,,
\end{align}
whereas the recursion relations are
\begin{align} \label{a11}
& (2\kappa+n+1) a_n  = (h-u_1)b_n- u_2b_{n-1} - u_3b_{n-2}\,,\\
& n b_n  = (u_1+h) a_{n-2}+ u_2 a_{n-3}+ u_3 a_{n-4}\,.
\end{align}

\section{One-potential Dirac Green function for a general potential}
\label{app:2}

For the evaluation of the self-energy correction, the
one-potential Dirac Green function $G^{(1)}$ is needed. Its radial part is
defined as
\begin{align} \label{b1}
G_{\kappa}^{(1)}(E,r_1,r_2) = \int_{0}^{\infty}dz\, z^2\,V(z)\, G^{(0)}_{\kappa}(E,r_1,z)
   \,G^{(0)}_{\kappa}(E,z,r_2)\,,
\end{align}
where $G^{(0)}$ is the free Dirac Green function. Substituting the
representation (\ref{gr01}) for $G^{(0)}_{\kappa}$ into (\ref{b1})
and introducing the integral
functions
\begin{align} \label{b2}
& J_{\kappa}^{00}(r) = \int_0^r dz\,z^2\,V(z)\,\varphi^{0^T}_{\kappa}(z)\, \varphi^0_{\kappa}(z)\,,\\
& J_{\kappa}^{i0}(r) = \int_0^r dz\,z^2\,V(z)\,\varphi^{{\infty}^T}_{\kappa}(z)\, \varphi^0_{\kappa}(z)\,,\\
& J_{\kappa}^{ii}(r) = \int_r^{\infty}
  dz\,z^2\,V(z)\,\varphi^{{\infty}^T}_{\kappa}(z)\, \varphi^{\infty}_{\kappa}(z)\,,
\end{align}
where $\varphi_{\kappa}^{(0)}$
$\left( \varphi_{\kappa}^{(\infty)}\right)$ denote the
regular (irregular) solutions of the free Dirac equation, we
write the one-potential Dirac Green function for $r_1\leq r_2$ as
\begin{align} \label{b3}
G_{\kappa}^{(1)}(E,r_1,r_2)
= \Phi^0_{\kappa}(r_1)\,\varphi_{\kappa}^{{\infty}^T}(r_2)+
   \varphi^0_{\kappa}(r_1)\,\Phi_{\kappa}^{{\infty}^T}(r_2) \,,
\end{align}
where
\begin{align} \label{b4}
&\ \Phi^0_{\kappa}(r) =  \varphi_{\kappa}^{\infty}(r)\,J_{\kappa}^{00}(r)-
               \varphi_{\kappa}^{0}(r)\,J_{\kappa}^{i0}(r) \,,\\
&\ \Phi^{\infty}_{\kappa}(r) =
  \varphi_{\kappa}^{\infty}(r)\,J_{\kappa}^{i0}(r)+
               \varphi_{\kappa}^{0}(r)\,J_{\kappa}^{ii}(r)\,.
\end{align}
For $r_1 >r_2$, the one-potential Green function is obtained by the symmetry condition,
\begin{align} \label{b5}
G_{\kappa}^{(1)}(E,r_1,r_2) = G_{\kappa}^{{(1)}^T}(E,r_2,r_1)\,.
\end{align}

Analogously to the approach used for the full Dirac Green function, we store
the functions $\varphi_{\kappa}$ and $\Phi_{\kappa}$ on a radial grid
$\left\{r_i\right\}_{i=1}^N$ and
obtain the one-potential Green function by interpolation. The integral functions
$J_{\kappa}$ are evaluated by numerical integration with help of
Gauss-Legendre quadratures. The integration interval $(0,\infty)$ is breaked
up at the position of the mesh points $r_i$, so that only one integral over
$(0,\infty)$ needs to be evaluated for a given value of $E$.
Analogously to the case of the full
Green function, all manipulations with the regular and irregular solutions are
carried out after normalizing them according to Eqs.~(\ref{gr03}) and
(\ref{gr03a}), in order to prevent numerical overflow and underflow.
Similar method of
computation of the one-potential Green function was used long ago by
M.~Gyulassy in his evaluation of the vacuum-polarization \cite{gyulassy:75}.


\end{document}